\documentclass[pra,aps,reprint,amsmath,amssymb,nofootinbib,superscriptaddress,10pt]{revtex4-1}
  
\usepackage{graphicx,dcolumn,bm}
\usepackage{amsmath}
\usepackage{amsthm}
\usepackage{amsfonts}
\usepackage{mathrsfs}
\usepackage[usenames,dvipsnames]{xcolor}
\usepackage[colorlinks=true,citecolor=MidnightBlue,linkcolor=MidnightBlue,urlcolor=MidnightBlue]{hyperref}
\usepackage{braket}
\usepackage{bbm}

\renewcommand{\vec}[1]{{\bf #1}}

\begin{document}
\title{Generation of N00N-like interferences with two thermal light sources}

\author{D.~Bhatti}\email[]{daniel.bhatti@fau.de}
\author{A.~Classen}
\author{R.~Schneider}
\affiliation{Institut f\"{u}r Optik, Information und Photonik, Friedrich-Alexander-Universit\"{a}t Erlangen-N\"{u}rnberg (FAU), 91058 Erlangen, Germany}
\affiliation{Erlangen Graduate School in Advanced Optical Technologies (SAOT), Friedrich-Alexander-Universit\"{a}t Erlangen-N\"{u}rnberg (FAU), 91052 Erlangen, Germany}
\author{S.~Oppel}
\affiliation{Institut f\"{u}r Optik, Information und Photonik, Friedrich-Alexander-Universit\"{a}t Erlangen-N\"{u}rnberg (FAU), 91058 Erlangen, Germany}
\author{J.~von~Zanthier}
\affiliation{Institut f\"{u}r Optik, Information und Photonik, Friedrich-Alexander-Universit\"{a}t Erlangen-N\"{u}rnberg (FAU), 91058 Erlangen, Germany}
\affiliation{Erlangen Graduate School in Advanced Optical Technologies (SAOT), Friedrich-Alexander-Universit\"{a}t Erlangen-N\"{u}rnberg (FAU), 91052 Erlangen, Germany}
%
%
%

\begin{abstract}
Measuring the $M$th-order intensity correlation function of light emitted by two statistically independent thermal light sources may display N00N-like interferences of arbitrary order $N = M/2$.
We 
show that via a particular choice of detector positions one can isolate $M$-photon quantum paths where either all $M$ photons are emitted from the same source or 
$M/2$ photons are collectively emitted by both sources.
The latter superposition displays N00N-like oscillations with $N = M/2$ which may serve, e.g., in astronomy, for imaging two distant thermal sources with $M/2$-fold increased resolution. We also discuss slightly modified detection schemes improving the visibility of the N00N-like  interference pattern and present measurements verifying the theoretical predictions.
\end{abstract}
%
%
\maketitle

\section{Introduction}
\label{intro}

N00N-states describe the collective propagation of $N$ identical particles in 
a two-path interferometer according to the superposition \cite{Boto(2000)}
\begin{equation}
 \ket{\text{N}00\text{N}} = \frac{1}{\sqrt{2}} \left( \ket{N,0} + e^{i N \varphi}\ket{0, N} \right) \, ,
\label{eq:N00N}
\end{equation}
where the single particle phase $\varphi$ 
is enhanced by the factor $N$  leading to an effective de Broglie wavelength of $\lambda/N$.
Given a N00N-state, the $N$ particle absorption rate $G^{(N)}({\mathbf r}, \ldots, {\mathbf r}) \propto 1 + \cos N \varphi$ exhibits a fringe spacing $N$ times as narrow as the one obtained
for a single particle in the same interferometer.

Since its first introduction in 1989 \cite{Sanders(1989)}, various aspects of  N00N-states have been explored, leading to numerous proposals and applications for quantum-enhanced measurements.
For example, superresolution \cite{Oppel(2012)}
and phase supersensitivity \cite{Israel(2014)}
have been studied in the context of quantum lithography \cite{Boto(2000),DAngelo(2001),Edamatsu(2002)}, quantum metrology \cite{Lee(2002),Mitchell(2004),Dowling(2008),Fogarty(2013)}, and quantum imaging 
\cite{Agafonov(2008),Agafonov(2009),Oppel(2012),Israel(2014)}. 
The interest in producing photonic N00N-states by superconducting systems \cite{Wang(2011),Strauch(2012),Su(2014),Xiong(2015),Chen(2017)} led recently to a proposal to generate double N00N-states \cite{Su(2017)}.
Also other quantum systems have been explored, leading to atomic N00N-states \cite{Hallwood(2010),Chen(2010),Schloss(2016),Song(2016)}, spin N00N-states \cite{Jones(2009)}, and 
even mechanical N00N states, implemented by entangling two mechanical micro-resonators \cite{Ren(2013),Macri(2016)}.
N00N-states have been further examined in the context of fundamental investigations of quantum mechanics \cite{Bergmann(2016),Teh(2016),Compagno(2017)}. 
Yet, both the realization and the detection of genuine N00N-states with a high particle number $N$ remains a challenge, being limited so far to a maximum of $N=9$ particles
\cite{Afek(2010),Israel(2012),Zhang(2016)arxiv}. 

In a different approach, 
it has been demonstrated that N00N-like modulations can also be produced by synthesizing patterns from sequential measurements by use of coherent light fields \cite{Resch(2007),Kothe(2010),Shabbir(2013)}.
It has further been shown that N00N-like fringe patterns can be obtained by measuring the $K$th-order intensity correlation function $G_{K \, \text{TLS}}^{(K)} ({\mathbf r}_1, \ldots, {\mathbf r}_K)$ of light (with $K = N + 1$) emitted 
by $K$ thermal light sources (TLS), positioned equidistantly along a line at distances $d \gg \lambda = 2 \pi/k$  \cite{Oppel(2012)} (see Fig.~\ref{fig:setup_2plusN}). 
Here, for $K-1$ detectors located at the so-called \textit{magic positions} 
\begin{equation}
	\delta_{j}=kd \sin\theta_{j}=2\pi \frac{j-2}{K-1} \ , \ j=2,\hdots,K \, ,
\label{eq:MagicPositionsN}
\end{equation}
the $K$th-order intensity correlation as a function of $\delta_{1}=kd \sin\theta_{1}$ takes the form 
\cite{Oppel(2012)}
\begin{equation}
	G_{K \, \text{TLS}}^{(K)}(\delta_1) \propto 1 + \mathcal{V}_{K-1} \cos((K-1) \delta_1) \, ,
\label{eq:N00NGN}
\end{equation}
with a visibility $\mathcal{V}_{K-1} < 1$.

In a follow-up investigation it was revealed that for an arbitrary arrangement of $K$ TLS on a grid with lattice constant $d$
and by placing $M-1$ detectors at the magic positions,
the $M$th-order intensity correlation function $G^{(M)}_{K \, \text{TLS}}(\delta_{1})$ as a function of $\delta_{1}$ oscillates only at selected spatial frequencies $f_i$ of the source arrangement, namely at those for which $f_i x_{1} = \kappa (M-1) \delta_1$, $\kappa \in \mathbb{Z}$ \cite{Classen(2016),Schneider(2018)}.
In this case, the $M$th-order intensity correlation function can take the form 
\begin{equation}
	G_{K \, \text{TLS}}^{(M)}(\delta_1) \propto 1 + \mathcal{V}_{\kappa (M-1)} \cos[\kappa (M-1) \delta_1] \, ,
\label{eq:N00NGm}
\end{equation}
with $\mathcal{V}_{\kappa (M-1)}< 1$.
From the above condition 
for the spatial frequencies $f_i$, it can readily be seen that the fastest modulation of $	G_{K \, \text{TLS}}^{(M)}(\delta_1)$ is given by 
$\kappa (M-1) = l_{max}$ (where $l_{max} d = (K-1) \, d$ denotes the separation of the two outer sources), identical to the frequency of Eq.~(\ref{eq:N00NGN}).

In this paper we show that by use of only two TLS
an arbitrarily fast sinusoidal oscillation of the $M$th-order intensity correlation function $G^{(M)}_{2 \, \text{TLS}}$ can be obtained. This happens when $M_{2}$ detectors are located at the magic positions and -- instead of a single moving detector -- $M_{1} \geq M_{2}$ moving detectors are employed, with $M = M_1 + M_2$.  In particular, we demonstrate that for two different detector configurations we can achieve N00N-like oscillations with $N=M_{2}$, for arbitrary $N \in \mathbb{N}$. This may serve, e.g., in astronomy, for imaging two distant thermal sources with $M_{2}$-fold increased resolution.

The paper is organized as follows: in Sect.~\ref{sec:MthOrder} we present the investigated setup and introduce the $M$th-order intensity correlation function in the far field of $K$ equidistantly aligned TLS, rewriting the correlation function in terms of the final quantum states of the TLS
\cite{Classen(2016)}. In Sect.~\ref{sec:Setup1} we show that by use of only two TLS and employing two sets of detectors, i.e., $M_{1}=M/2$ moving detectors plus $M_{2}=M/2$ fixed detectors, the $M$th-order intensity correlation function displays N00N-like oscillations of arbitrary order $N=M/2$.
We explain this behavior in a quantum path picture demonstrating  that only those $M$-photon quantum paths contribute to the interference signal for which the detected photons are emitted in a N00N-like manner.
In Sect.~\ref{sec:Setup2} we discuss a slightly modified setup, employing $M_{2}$ fixed detectors and $M_{1} \geq M_{2}$ moving detectors with $M = M_1 + M_2$, to produce identical interference patterns, however, with an increased visibility. In Sect.~\ref{sec:ThermalN00NState} we investigate  the projected density matrix of the two TLS after the first $M_2$ photons have been recorded and compare it to the quantum mechanical N00N-state with $N=M_2$. Experimental results of N00N-like interferences for the setup featured in Sect.~\ref{sec:Setup2} are presented in Sect.~\ref{sec:Experiment} and the increase in resolution is discussed in Sect.~\ref{sec:Discussion}. We finally present our conclusions in Sect.~\ref{sec:Conclusion}.

\section{\texorpdfstring{$M$th}{Mth}-Order Intensity Correlation Function}
\label{sec:MthOrder}

\begin{figure}[t]
	\centering
		\includegraphics[width=0.8\columnwidth]{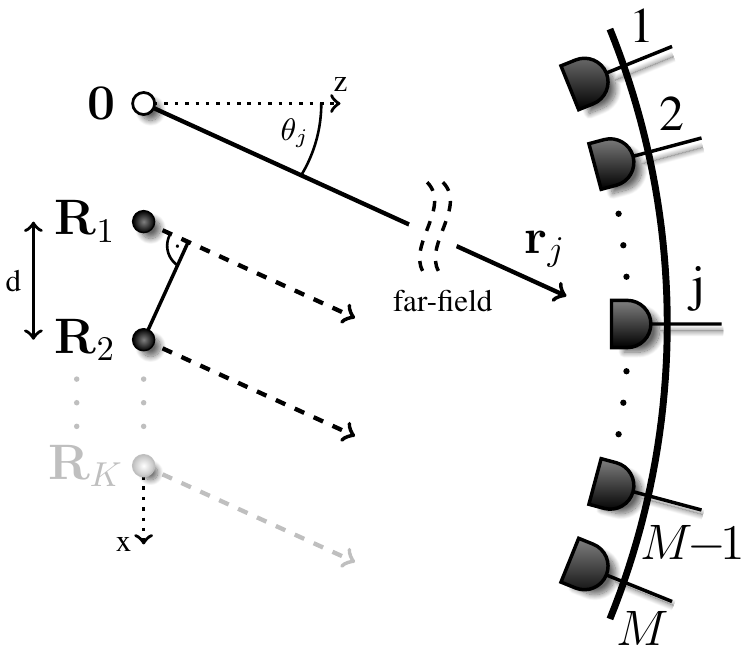}
	\caption{Considered setup: Two thermal light sources (TLS) are located along the $x$ axis at the positions $\vec{R}_{1}$ and $\vec{R}_{2}$ with a separation $d\gg \lambda$. To measure 
the $M$th-order intensity correlation function $G^{(M)}_{2}(\vec{r}_{1},\hdots, \vec{r}_{M})$, $M$ detectors are placed in the far field of the sources at positions $\vec{r}_{j}$, $j=1,\hdots,M$. 
As indicated in the figure, the setup can easily be extended to an arbitrary number of sources $K$ located along the $x$ axis at the positions $\vec{R}_{3}, \ldots, \vec{R}_{K}$.}
	\label{fig:setup_2plusN}
\end{figure}

For $K$ light sources, the coincident spatial $M$th-order intensity correlation function is defined as \cite{Glauber(1963)Quantum}
\begin{equation}
\begin{aligned}
	G^{(M)}_{K}(\vec{r}_{1},\hdots, \vec{r}_{M}) = \left<: \prod_{j=1}^{M} \hat{E}^{(-)}(\vec{r}_{j}) \hat{E}^{(+)}(\vec{r}_{j}) : \right>_{\rho_{{}_K}} ,
\label{eq:DefGN}
\end{aligned}
\end{equation}
where $\left< : \hat{\mathcal{A}} :\right>_{\rho_{{}_K}}$ denotes the (normally ordered) quantum mechanical expectation value of an operator $\hat{\mathcal{A}}$ for a field in the state $\rho_{{}_K}$. In Eq.~(\ref{eq:DefGN}), the positive and negative frequency part of the electric field operators at position $\vec{r}_{j}, j=1,\hdots,M$, in the far field of the sources, $\hat{E}^{(+)}(\vec{r}_{j})$ and $\hat{E}^{(-)}(\vec{r}_{j})$, respectively,
are given by \cite{Classen(2016)}
\begin{equation}
\begin{aligned}
	\left[ \hat{E}^{(-)}(\vec{r}_j) \right]^\dagger = \hat{E}^{(+)}(\vec{r}_j) \propto \sum_{l=1}^{K} e^{-i\,\alpha_{l} \delta_{j}} \;\hat{a}_{l} ,
\label{eq:DefE2}
\end{aligned}
\end{equation}
where $\hat{a}_{l}$ describes the bosonic annihilation operator for a photon emitted from source $l$.
We assume the sources to be equidistantly arranged along the $x$ axis at the positions $\vec{R}_{l}=l\vec{R}_{1}=l(d,0,0), l=1,\hdots,K$ (see Fig.~\ref{fig:setup_2plusN}), with an intersource distance $d \gg \lambda$ such that any interaction between the sources can be neglected.
The optical phase $\alpha_{l} \delta_{j}$ accumulated by a photon emitted from source $l$ at $\vec{R}_{l}$ and detected at $\vec{r}_{j}$ relative to a photon emitted at $\vec{R}_{1}$  is given in the far-field limit by
\begin{equation}
	\alpha_{l} \delta_{j} = k \frac{(\vec{R}_{l}-\vec{R}_{1})\cdot\vec{r}_{j}}{r_{j}} = \alpha_{l}kd \sin\theta_{j} \ ,
\label{eq:DefDelta}
\end{equation}
where $\alpha_l$ corresponds to the relative distances between the first and the $l$th source in units of $d$, i.e., $\alpha_{l}=l-1$ (see Table~\ref{tab:N2}).

Plugging Eq.~(\ref{eq:DefE2}) into Eq.~(\ref{eq:DefGN}) 
the $M$th-order correlation function calculates to \cite{Classen(2016)}
\begin{align}
  \label{eq:GN}
	 & G^{(M)}_{K \,\text{TLS}}(\delta_1,\ldots,\delta_M) \nonumber \\[2mm]
  &  = \langle\hat{E}^{(-)}(\delta_1)\ldots\hat{E}^{(-)}(\delta_M)\hat{E}^{(+)}(\delta_M)\ldots\hat{E}^{(+)}(\delta_1)\rangle_{\rho_{{ }_{K \,\text{TLS}}}} \nonumber \\[2mm]
	& = \sum_{\{n_{l}\}}X_{\{n_{l}\}}|\sum_{\mathcal{P}_{\{\alpha_{l}\}}}e^{i(\alpha_{l_{1}}\delta_{1}+\alpha_{l_{2}}\delta_{2}+\cdots+\alpha_{l_{M}}\delta_{M})}|^{2} \, ,
\end{align}
where $\sum_{\{n_l \}}$ runs over all possible $M$-photon distributions $\{n_l \}=\{ n_1, \ldots , n_{K} \}$ among the $K$ sources, i.e., over all partitionings of the $M$ photons when assigned to the emitting sources such that $\sum_{l=1}^{K}n_{l} = M$. Here, each partitioning $\{n_{l}\}$ describes a certain final state, whereas $X_{\{n_{l}\}}$ denotes the statistical loading according to the light statistics of the light field $\rho_{{ }_{K \,\text{TLS}}}$. In Eq.~(\ref{eq:GN}), $\mathcal{P}_{\{\alpha_{l}\}}$ is the permutation over all phase prefactors $\alpha_l$
representing all different yet indistinguishable $M$-photon quantum paths that result from a specific final state $\{n_{l}\}$ \cite{Classen(2016),Bhatti(2016)}.
Note that each thermal source can emit an arbitrary number of photons. Consequently, the complete set of phase prefactors reads $\{\alpha_l \}=\{\alpha_1,\hdots,\alpha_1,\alpha_2,\hdots,\alpha_{K},\hdots,\alpha_{K}\}$, where each phase prefactor $\alpha_l$ ($l=1,\hdots, K$) is contained $n_{l}$ times, what corresponds to $n_{l}$ photons being emitted by the $l$th source (see Table~\ref{tab:N2}). 

\renewcommand{\arraystretch}{1.4} 
\begin{table} [ht]
\begin{center}
		\begin{tabular}{|c|ccccc|}
		\hline
			source number $l$ & \hspace{1mm} \ 1 \ & \ 2 \ &  \ 3 \ &\ $\hdots$ \ & \ $K$ \\
			\hline
			phase prefactor $\alpha_{l}$ & \hspace{1mm} 0 & 1 & 2 &  $\hdots$ &  $K-1$ \\
			\hline
			$\#$ photons: final state & \hspace{1mm} $n_1$ & $n_2$ & $n_3$ &  $\hdots$ &  $n_{K}$ \\
			\hline
\end{tabular}
\end{center}
\vspace{-3mm}
\caption{Source numbers $l=1,\hdots, K$ with corresponding relative phase prefactors $\alpha_{l}$ and number of emitted photons for a given final state $\{n_l \}=\{ n_1, \ldots , n_{K} \}$.}
\label{tab:N2}
\end{table}
\renewcommand{\arraystretch}{1.0} 

\section{Setup 1: \texorpdfstring{$M/2$}{M/2} fixed detectors at magic positions, \texorpdfstring{$M/2$}{M/2} detectors at moving magic positions}
\label{sec:Setup1}

In the following we focus on the special case of two TLS, placed at the positions $\vec{R}_{1}$ and $\vec{R}_{2}$ with $d = |\vec{R}_{2} - \vec{R}_{1}| \gg \lambda$, and investigate the $M$th-order intensity correlation function for an even number of detectors $M$.
Out of the $M$ detectors, we assume that $M/2$ detectors are placed at the fixed magic positions (MP) [see Eq.~(\ref{eq:MagicPositionsN})] 
\begin{equation}
	\delta_{j}=2\pi \frac{j-2}{M/2} \ , \ j=2,\hdots,M/2+1 \, ,
\label{eq:MagicPositionsM}
\end{equation}
while a second set of $M/2$  detectors are placed at the so-called \textit{moving magic positions} (MMP)
\begin{equation}
	\delta_{1,j}=\delta_{1} + 2\pi \frac{j-2}{M/2} \ , \ j=2,\hdots,M/2 + 1 \, .
\label{eq:MovingMagicPositions}
\end{equation}
In this case, the  $M$th-order correlation function
calculates to [cf. Eq.~(\ref{eq:GN})]
\begin{align}
  \label{eq:GNmovingMP}
	  G^{(M)}_{2 \,\text{TLS}}\textbf{(}&\text{MMP}(\delta_{1});\text{MP}\textbf{)}  \nonumber \\
		=& \sum_{\{n_{l}\}}X_{\{n_{l}\}}\Big| \sum_{\substack{ \{\beta_{k}\} + \{\gamma_{k}\} = \{\alpha_{k}\} \\[0.25mm]  |  \{\beta_{k}\} | =| \{\gamma_{k}\} | =M/2}}  \nonumber \\
		& \times \sum_{\mathcal{P}_{\{\beta_{k}\}}} e^{i(\beta_{k_{1}}\delta_{1,2}+\cdots+\beta_{k_{M/2}}\delta_{1,M/2+1})} \nonumber \\
		& \times \sum_{\mathcal{P}_{\{\gamma_{k}\}} } e^{i(\gamma_{k_{1}}\delta_{2} + \cdots + \gamma_{k_{M/2}}\delta_{M/2+1})} \Big|^{2} \, ,
\end{align}
where the sum over all $M$-photon quantum paths $\sum_{\mathcal{P}_{\{\alpha_{l}\}}}$ of Eq.~(\ref{eq:GN}) has been restructured by first dividing the $M$ photons into two subgroups of $M/2$ photons, each with phase prefactors $ \{\beta_{k}\}$ and $\{\gamma_{k}\}$ and cardinality $ | \{\beta_{k}\}| = |\{\gamma_{k}\}| = M/2$ [see the sum in line 2 of Eq.~(\ref{eq:GNmovingMP})], and then permuting these new groups of phase prefactors to consider all $M$-photon quantum paths. Note that for a particular final state $\{ n_{l} \}$ the sum $\sum_{\mathcal{P}_{\{\beta_{k}\}}}$ in line 3 of Eq.~(\ref{eq:GNmovingMP}) denotes all possibilities of $M/2$-photons to be detected at the MMP, while the sum $\sum_{\mathcal{P}_{\{\gamma_{k}\}}}$ in line 4 of Eq.~(\ref{eq:GNmovingMP}) denotes all possibilities of $M/2$-photons to be recorded at the MP.

In Ref.~\cite{Classen(2016)} it has been shown that in line 4 of Eq.~(\ref{eq:GNmovingMP}) the particular sum over all permutations of $M/2$ phase prefactors $\mathcal{P}_{\{\gamma_{k}\}}$ 
with $M/2$ detectors at the MP can only be unequal to zero if $\sum_{j=1}^{M/2} \gamma_{k_{j}}= 0$, mod$(M/2)$. This can only be fulfilled if all $M/2$ photons appearing in the sum are emitted from the same source, either source $1$ or source $2$, i.e., all $\gamma_{k_{j}}$
either take the value $\gamma_{k_{j}}=0$ or $\gamma_{k_{j}}=1$.
All other $M/2$-photon quantum paths vanish. 
In the case that all $M/2$ photons appearing in the sum are emitted from source 1, i.e., $\gamma_{k_{j}}=0$, 
the sum yields
\begin{align}
  \label{eq:sum_detectorphases0}
&\sum_{\mathcal{P}_{\{\gamma_{k}\} = \{0,\hdots,0\}}} \text{exp}\left(0\right) = 1 ,
\end{align}
whereas in the other case that all $M/2$ photons
are emitted from source 2, i.e., $\gamma_{k_{j}}=1$,
the sum yields $\pm1$, depending on the number of detectors $M/2$ being even or odd
\begin{align}
  \label{eq:sum_detectorphases}
\sum_{\mathcal{P}_{\{\gamma_{k}\} = \{1,\hdots,1\}}} &\text{exp}\left(i {\sum_{j=2}^{M/2+1}\delta_{j}}\right) \nonumber \\[1.5mm]
=\ &  \text{exp}\left(\frac{i2\pi}{M/2}\sum_{j=2}^{M/2+1} (j-2)\right) \nonumber \\[1.5mm]
=\ &  \text{exp}\left(\frac{i2\pi}{M/2} \frac{(M/2-1)M/2}{2} \right) \nonumber \\[1.5mm]
=\ &  \text{exp}\left[ i (M/2-1)\pi \right] .
\end{align}
\begin{figure}[b]
	\centering
		\includegraphics[width=1.0\columnwidth]{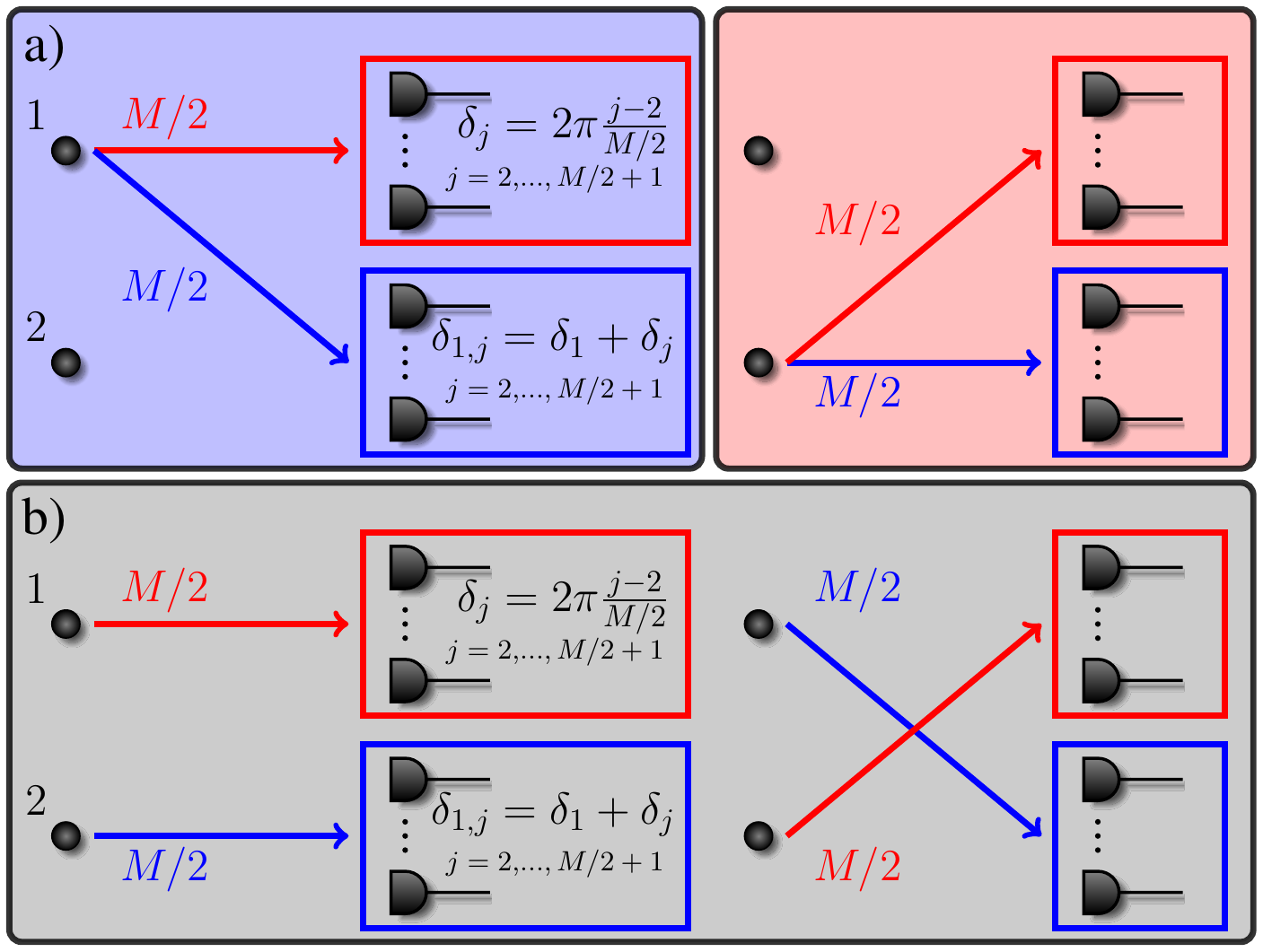}
	\caption{Setup 1: Quantum paths that contribute to the $M$th-order correlation function and lead to N$00$N-like modulations for one set of $M/2$ fixed detectors at the magic positions (MP) [see Eq.~(\ref{eq:MagicPositionsM})], while a second set of $M/2$ detectors is placed at the moving magic positions (MMP) [see Eq.~(\ref{eq:MovingMagicPositions})].}
	\label{fig:N00N_quantum_paths_setup1}
\end{figure}

To calculate in line 3 of Eq.~(\ref{eq:GNmovingMP}) the remaining sum 
over all permutations of phase prefactors $\{\beta_{k} \}$ with $M/2$ detectors at the MMP, it is helpful to factorize the global phase $(\beta_{k_{1}}+\cdots +\beta_{k_{M/2}})\delta_{1}$, so that the sum becomes identical to the already discussed sum $\sum_{\{ \gamma_{k}\}}$ [with identical results of Eqs.~(\ref{eq:sum_detectorphases0}) and (\ref{eq:sum_detectorphases})].
This means that for the set of moving detectors $\delta_{1,j}$,
the same conditions hold as for the set of fixed detectors $\delta_{j}$,
i.e., for a non-vanishing contribution either the first source or the second source has emitted all $M/2$ photons, detected at the MP or the MMP, respectively.
Considering now all $M$ recorded photons, it follows that only two partitions can lead to a valid measurement event: a) all $M$ photons are emitted from the same source, i.e., either all photons from source 1 or all photons from source 2 (see Fig.~\ref{fig:N00N_quantum_paths_setup1}a), or b) $M/2$ photons are emitted from source 1 and $M/2$ photons from source 2  
(see Fig.~\ref{fig:N00N_quantum_paths_setup1}b).
Altogether the partitions in a) and b) describe four distinct $M$-photon quantum paths that contribute to the measured $M$th-order correlation signal. Here, the two possibilities from a) belong to two distinguishable final states and thus have to be added incoherently, whereas the two different yet indistinguishable possibilities from b) belong to a single identical final state and thus have to be added coherently.

In the simplest case of $M=2$ detectors, one fixed and one moving, the resulting modulations are well known from the landmark Hanbury Brown and Twiss (HBT) experiment
\cite{HBT(1956)Correlation,Brown(1967),HANBURY(1974)}, i.e., stemming from the interference of two different yet indistinguishable 2-photon quantum paths \cite{Fano(1961),Liu(2009),Bhatti(2016)}.
For $M > 2$, however, the predicted $M$-photon quantum path interferences
lead to N00N-like modulations with $N=M/2$. This will be discussed in the following.

By use of Eqs.~(\ref{eq:sum_detectorphases0}) and (\ref{eq:sum_detectorphases}), the $M$th-order correlation function of Eq.~(\ref{eq:GNmovingMP}) calculates to
\begin{align}
  \label{eq:GNmovingMPfinal}
	 & G^{(M)}_{2 \,\text{TLS}}\textbf{(}\text{MMP}(\delta_{1});\text{MP}\textbf{)} \nonumber \\[2mm]
	& = 2\, M! + [(M/2)!]^{2} \Big| (-1)^{M/2-1} + (-1)^{M/2-1} e^{i (M/2) \delta_{1}} \Big|^{2} \nonumber \\
	& = 2\, [(M/2)!]^{2} \left\{\binom{M}{M/2} + 1 + \cos[(M/2)\delta_{1}] \right\}.
\end{align}

Again, setting $M=2$ in Eq.~(\ref{eq:GNmovingMPfinal}), we find for the interference term the well-known modulation $\cos(\delta_{1})$ with a visibility of $\mathcal{V}_{2} = 1/3$, as in the original HBT experiment \cite{HBT(1956)Correlation,Brown(1967),HANBURY(1974)}. 
However, for $M>2$, we obtain a modulation $\cos[(M/2) \delta_{1}]$, being $M/2$ times as fast as the modulation of the HBT experiment, i.e., a fringe pattern of identical shape as a N00N-type oscillation with $N=M/2$ [cf. Eq.~(\ref{eq:N00N})]. 

Note that in contrast to the results of Refs.~\cite{Oppel(2012),Classen(2016)}, where the sinusoidal modulation $\sim K-1 = N$ of the $K$th-order correlation function is produced by $K$ TLS [cf. Eq.~(\ref{eq:N00NGN})], we obtain here a modulation of arbitrary order $M/2=N$ by using only \textit{two} TLS. The difference arises, since in Refs.~\cite{Oppel(2012),Classen(2016)} projective measurements of photons at the MP are used to isolate within the $K$th-order correlation function only those interference terms appearing between the two outer sources, separated by a distance $(K-1)d$.  
In contrast, in the present setup, the measurement of photons at the MP \textit{and} the MMP are used to isolate within the $M$th-order correlation function those interference terms 
for which each of the two sources emits $M/2$ photons collectively propagating to one of the two detector sets (see Fig.~\ref{fig:N00N_quantum_paths_setup1}b).
Note that the corresponding quantum state producing the identical interference in the $M$th-order correlation function (yet with a visibility of $100\%$) by use of the same detection scheme is the twin fock state $\ket{M/2,M/2}$ \cite{Dowling(2008)}. Indeed if we plugged this quantum state into Eq.~(\ref{eq:GNmovingMP}), we would exclusively obtain the quantum paths of Fig.~\ref{fig:N00N_quantum_paths_setup1}b.


In comparison to the quantum mechanical N00N-states \cite{Sanders(1989),Boto(2000)}, the visibility $\mathcal{V}_{M}$ of the N00N-like interferences obtained in the present setup is reduced. This is due to the contributing partitions where all $M$ photons are emitted from the same source (see Fig.~\ref{fig:N00N_quantum_paths_setup1}a). According to Eq.~(\ref{eq:GNmovingMPfinal}), the visibility is given by
\begin{equation}
	\mathcal{V}_{M} = \frac{[(M/2)!]^{2}}{[(M/2)!]^{2}+M!} \ .
\label{eq:visibility2(m-1)}
\end{equation}
%
Explicitly, up to the orders $M= 2, \ldots, 10$, $\mathcal{V}_{M}$ calculates to $\mathcal{V}_{2}\approx0.33$, $\mathcal{V}_{4}\approx 0.1429$, $\mathcal{V}_{6}\approx 0.0476$, $\mathcal{V}_{8}\approx 0.0141$, $\mathcal{V}_{10}\approx 0.0040$.

\section{Setup 2: \texorpdfstring{$M_2$}{M\string_{2}} fixed detectors at magic positions, \texorpdfstring{$M_1$}{M\string_{1}} moving detectors at the same position}
\label{sec:Setup2}

Next, let us investigate the configuration where out of the $M$ detectors used to measure the $M$th-order intensity correlation function $M_1$ detectors are located at the same position $\delta_{1}$ and $M_2 \leq M_1$ detectors are placed at the MP, with $M = M_1 + M_2$. 
This configuration can be used to increase the visibility of the $M$th-order correlation function.

\begin{figure}[b]
	\centering
		\includegraphics[width=1.0\columnwidth]{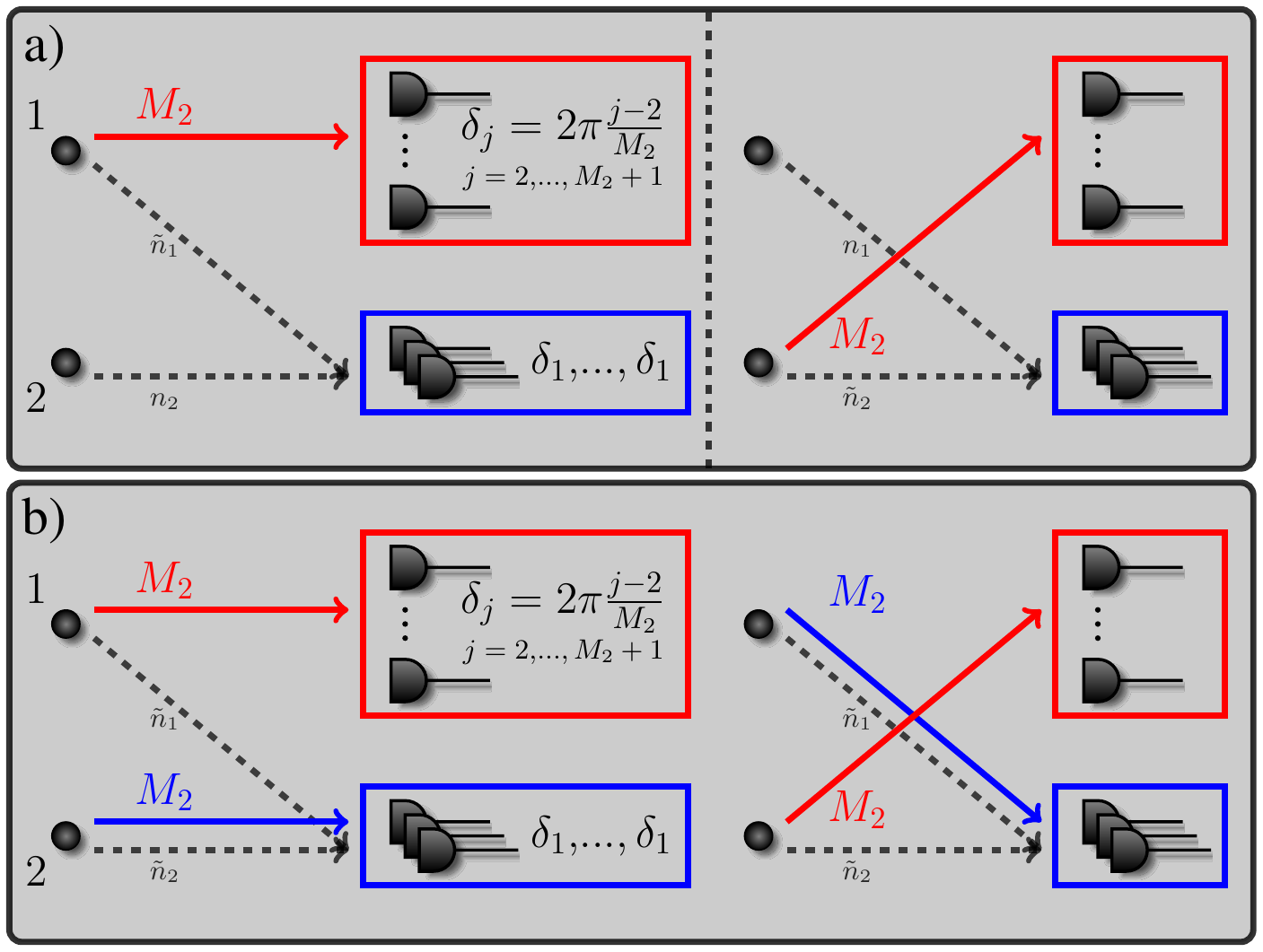}
	\caption{Setup 2: a) Possible $M$-photon quantum paths for $M_{2}$ fixed detectors at the magic positions (MP) $\delta_{j}$ ($j=2,\hdots,M_{2}+1$) and $M_{1}$ moving detectors at the same position $\delta_{1}$. The $M_{2}$ photons detected at the MP either have to be emitted completely from source 1 (left part) or source 2 (right part). Note that the different quantum paths do not necessarily belong to the same quantum state. b) For $M_{1}\geq M_{2}$ interference occurs if source 1 and source 2 emit at least $M_{2}$ photons each. Out of the complete set of $M$-photon quantum paths  only the interfering $M$-photon quantum paths are shown leading to N$00$N-like modulations.}
	\label{fig:N00N_quantum_paths_setup2}
\end{figure}

To see this in more detail, we rewrite in Eq.~(\ref{eq:GN}) in the third line the absolute square in terms of the fixed detector phases $\delta_{2},\hdots,\delta_{M_2+1}$. We thus obtain
\begin{align}
  \label{eq:GNrewrite}
	  G^{(M)}_{2 \,\text{TLS}}&\textbf{(}\delta_{1},\ldots ,\delta_{1};\text{MP}\textbf{)} \nonumber \\
		&= \sum_{\{n_{l}\}}X_{\{n_{l}\}}\Big| \sum_{\substack{ \{\beta_{k}\} + \{\gamma_{k}\} = \{\alpha_{k}\} \\[0.25mm]  |  \{\beta_{k}\} | = M_{1},\, | \{\gamma_{k}\} | =M_{2}}}  \nonumber \\
			&\phantom{{}={}} \times \sum_{\mathcal{P}_{\{\beta_{k}\}}} e^{i(\beta_{k_{1}}+\cdots+\beta_{k_{M_{1}}})\delta_{1}} \nonumber \\
		&\phantom{{}={}} \times \sum_{\mathcal{P}_{\{\gamma_{k}\}} } e^{i(\gamma_{k_{1}}\delta_{2} + \cdots + \gamma_{k_{M_{2}}}\delta_{M_{2}+1})} \Big|^{2} \, ,
\end{align}
Analogous to the argumentation of Sect.~\ref{sec:Setup1}, we can make use of the fact that the sum over all permutations $\mathcal{P}_{\{\gamma_{k}\}}$ with $M_2$ detectors at the MP can only be unequal to zero if $\sum_{j=1}^{M_2} \gamma_{k_{j}}= 0$, mod$(M_2)$, i.e., all $\gamma_{k_{j}}$ either take the value $\gamma_{k_{j}}=0$ or $\gamma_{k_{j}}=1$, with $j=1,\hdots,M_2$ \cite{Classen(2016)}. Again, this means that all $M_{2}$ photons recorded simultaneously by the detectors at the MP have to be emitted by the same source (see Fig.~\ref{fig:N00N_quantum_paths_setup2}a). If all photons are emitted by source 1, i.e., $\gamma_{k_{j}}=0$,
the sum yields $+1$, whereas if all photons are emitted by source 2, i.e., $\gamma_{k_{j}}=1$,
the sum yields $\pm1$, depending on the number of detectors $M_2$ [see Eq.~(\ref{eq:sum_detectorphases0}) and (\ref{eq:sum_detectorphases}), respectively].

Making use of the information about the $M_2$ fixed detectors placed at the MP, Eq.~(\ref{eq:GNrewrite}) can thus be rewritten in the form
\begin{align}
  \label{eq:GNrewritemagic}
	 & G^{(M)}_{2 \,\text{TLS}}\textbf{(}\delta_{1},\ldots ,\delta_{1};\text{MP}\textbf{)} \nonumber \\
	& = \sum_{n_{1}+n_{2}=M}n_{1}!\, n_{2}! \Big| \binom{M_{1}}{n_{1}-M_2, n_{2}} \nonumber  \\
	& \phantom{{}={}} + (-1)^{M_2-1} \binom{M_{1}}{n_{1}, n_{2}-M_2} e^{- i M_2 \delta_{1}} \Big|^{2} \, ,
\end{align}
where the phase prefactors $\alpha_{1}=0$ and $\alpha_{2}=1$ have been employed and $\sum_{n_{1}+n_{2}=M}$ denotes the sum over all possibilities to distribute the total number of $M$ photons among the two sources (see Table~\ref{tab:N2}). From Eq.~(\ref{eq:GNrewritemagic}) it can be seen that interferences arise only if $n_{1}\geq M_{2}$ and $n_{2}\geq M_{2}$, while other $M$-photon quantum paths yield a constant contribution. The interfering $M$-photon quantum paths are depicted in Fig.~\ref{fig:N00N_quantum_paths_setup2}b.
Expanding the absolute square in Eq.~(\ref{eq:GNrewritemagic}), we finally obtain
\begin{align}
\label{eq:GNgeneralform}
	 & G^{(M)}_{2 \,\text{TLS}}\textbf{(}\delta_{1},\ldots ,\delta_{1};\text{MP}\textbf{)}  = C_{1} + (-1)^{M_2-1} C_{2} \cos(M_2\delta_{1}) ,
\end{align}
where the constant term $C_{1}$ is given by
\begin{align}
\label{eq:GNconstant}
	C_{1} = & \sum_{n_{1}+n_{2}=M} n_{1}!\, n_{2}! \left[ \binom{M_{1}}{n_{1}-M_2, n_{2}}^{2} \right. \nonumber  \\
	& \left. +  \binom{M_{1}}{n_{1}, n_{2}-M_2}^{2}  \right]  \nonumber \\
	 = & \,  2\, M_{1}! \, M_2! \sum_{k=0}^{M_{1}} \binom{M_{1}}{k} \binom{M_2+k}{k} ,
\end{align}
and $C_{2}$ describing the prefactor of the interference term reads
\begin{align}
\label{eq:GNinterference}
	C_{2} = & 2\, M_{1}! \sum_{n_{1}+n_{2}=M}  \binom{M_{1}}{n_{1}-M_2,n_{2}-M_2}  \nonumber \\[2mm]
	=&\frac{2(M_{1}!)^{2}}{(M_{1}-M_2)!} \sum_{\tilde{n}_{1}+\tilde{n}_{2}=M_{1}-M_2}  \binom{M_{1}-M_2}{\tilde{n}_{1},\tilde{n}_{2}}  \nonumber \\
	=& 2^{M_{1}-M_2+1}\frac{(M_{1}!)^{2}}{(M_{1}-M_2)!} \ .
\end{align}

The fact that interferences arise only if $n_{1}\geq M_2$ and $n_{2}\geq M_2$ implies that $M_{1}\geq M_2$ moving detectors are required -- in addition to the $M_2$ fixed detectors at the MP -- to produce a modulation of order $M_2$. In fact, as shown in Fig.~\ref{fig:N00N_quantum_paths_setup2}b, the interferences effectively result from $(2 M_2)$-photon quantum paths, where $M_2$ photons are emitted from source 1 and $M_2$ photons are emitted from source 2, similar as in setup~$1$ presented in Sect. \ref{sec:Setup1}, with $M_1=M_2$.
However, in setup~$2$ an additional number of photons $M_1-M_2$ has to be recorded by the detectors placed at $\delta_{1}$, which can be emitted by source 1 or source 2, without any further restriction. Thereby, new indistinguishable $M$-photon quantum paths arise which add to the same interference pattern, where the visibility is increased with every additional detector. 
This allows for setup 2 to exceed the visibility of setup 1, where the latter is given by Eq.~(\ref{eq:visibility2(m-1)}).
%

In the case of $M_2=2$ detectors at the MP and $M_{1}$ moving detectors at $\delta_{1}$ it is possible to calculate the $M$th-order correlation function $G_{2\,\text{TLS}}^{(2+M_{1})}$ of Eq.~(\ref{eq:GNgeneralform}) in an explicit form. One obtains
\begin{align}
\label{eq:GNgeneralform_2}
	&G_{2\,\text{TLS}}^{(2+M_{1})}(\delta_{1},\hdots,\delta_{1};0,\pi) \nonumber \\
	&= 2^{M_{1}-1} M_{1}! \left[M_{1}^{2} + 7M_{1} +  8 - M_{1}(M_{1}-1) \cos(2\delta_{1}) \right] \, ,
\end{align}
with a visibility given by
\begin{align}
\label{eq:GNgeneralform_2_vis}
	\mathcal{V}_{2+M_{1}}=\frac{M_{1}(M_{1}-1)}{M_{1}^{2} +7M_{1}+ 8} .
\end{align}
Comparing $\mathcal{V}_{2+M_{1}}$ for $M_{1}$ moving detectors at the same position $\delta_{1}$ to the visibility $\mathcal{V}_{4}=1/7$ of setup 1 [cf. Eq~(\ref{eq:visibility2(m-1)})], where 2 detectors are placed at the MMP $\delta_{1}$ and $\delta_{1}+\pi$, respectively, we see that  $\mathcal{V}_{2+M_{1}}> \mathcal{V}_{4}$ already for $M_{1}\geq 3$ moving detectors.

For higher numbers of detectors, i.e., $M_2>2$, the $M$th-order correlation function can not be written in a simple analytic form [see Eqs.~(\ref{eq:GNgeneralform})-(\ref{eq:GNinterference})]. However, it can be calculated how many moving detectors are required for setup 2 to exceed the visibilities of setup 1: In the cases of $M_2=2,3,4,5$ fixed detectors one would have to use at least $M_{1}=3,5,6,7$ moving detectors at the same position, respectively.

\section{N00N-like states for classical sources}
\label{sec:ThermalN00NState}

In previous work we have identified an isomorphism between $G_{\rho_{{}_K}}^{(M)}$ and $G_{\tilde{\rho}_{{}_K}^{(M-1)}}^{(1)}$, where, starting from a light field $\rho_{{}_K}$ generated by $K$ quantum or classical light sources, the $M$th-order correlation function can be written as \cite{Bhatti(2016),Wiegner(2015),Bhatti(2018)}
\begin{align}
	G_{\rho_{{}_K}}^{(M)}(\vec{r}_{1},\ldots,\vec{r}_{M}) = G_{\tilde{\rho}_{{}_K}^{(M-1)}}^{(1)}(\vec{r}_{1}) G_{\rho_{{}_K}}^{(M-1)}(\vec{r}_{2},\ldots,\vec{r}_{M})  \, ,
\label{eq:Isomorphism}
\end{align}
where $\tilde{\rho}_{{}_K}^{(M-1)}$ denotes the state after $M-1$ photons have been recorded at the positions $\vec{r}_{2},\ldots,\vec{r}_{M}$. In general, this state can be expressed as 
\begin{align}
	\tilde{\rho}_{{}_K}^{(M-1)} = \frac{\left[ \prod_{j=2}^{M} \hat{E}^{(+)}(\vec{r}_{j}) \right] \rho_{{}_K} \left[ \prod_{j=2}^{M} \hat{E}^{(-)}(\vec{r}_{j}) \right] }{G_{\rho_{{}_K}}^{(M-1)}(\vec{r}_{2},\ldots,\vec{r}_{M})}  \, .
\end{align}
For the two setups presented in Sects.~\ref{sec:Setup1} and \ref{sec:Setup2}, we adjust the isomorphism and employ it for the first $M_{2}$ photon detections at the MP and the subsequent $M_{1}$ photon detections at the moving detectors (MD). Note that in setup 1 we have $M_{1}=M_{2}=M/2$ at the MP and the MMP, respectively, while in setup 2 we have $M_{2}$ detectors at the MP, and $M_{1}\geq M_{2}$ detectors at the moving position $\delta_{1}$. Independent of the setup we can thus write
\begin{align}
	G_{\rho_{{}_K}}^{(M)}(\text{MD}(\delta_{1});\text{MP}) = G_{\tilde{\rho}_{{}_K}^{(M_{2})}}^{(M_{1})}(\text{MD}(\delta_{1})) G_{\rho_{{}_K}}^{(M_{2})}(\text{MP})  \, ,
\label{eq:IsomorphismMP}
\end{align}
where the state $\tilde{\rho}_{{}_K}^{(M_{2})}$ after the detection of the first $M_{2}$ photons at the MP takes the form
\begin{align}
	\tilde{\rho}_{{}_K}^{(M_{2})} = \frac{\left[ \prod_{j=2}^{M_{2}+1} \hat{E}^{(+)}(\delta_{j}) \right] \rho_{{}_K} \left[ \prod_{j=2}^{M_{2}+1} \hat{E}^{(-)}(\delta_{j}) \right] }{G_{\rho_{{}_K}}^{(M_{2})}(\text{MP})}  \, .
\label{eq:StateMP}
\end{align}
By using the definitions for $\hat{E}^{(+)}(\delta_{j})$ and $\hat{E}^{(-)}(\delta_{j})$ of Eq.~(\ref{eq:DefE2}) for $K=2$ TLS one can calculate the products in Eq.~(\ref{eq:StateMP}) explicitly to obtain 
\begin{align}
	&\left[\prod_{j=2}^{M_{2}+1} \hat{E}^{(-)}(\delta_{j})\right]^{\dagger} \nonumber \\
	&= \prod_{j=2}^{M_{2}+1} \hat{E}^{(+)}(\delta_{j}) \nonumber \\
	&= \prod_{j=2}^{M_{2}+1} \left( e^{-i\alpha_{1} \delta_{j}} \hat{a}_{1} + e^{-i \alpha_{2} \delta_{j}}\hat{a}_{2} \right) \nonumber \\
	&	= \sum_{m_{1}+ m_{2} = M_{2}} \hat{a}_{1}^{m_{1}}\hat{a}_{2}^{m_{2}} \sum_{\mathcal{P}_{\{\alpha_{k}\}}} e^{-i(\alpha_{k_{1}}\delta_{2} + \cdots + \alpha_{k_{M_{2}}}\delta_{M_{2}+1})} \, ,
\label{eq:FieldMP}
\end{align}
where the first sum in the last line of Eq.~(\ref{eq:FieldMP}) runs over all possibilities of $m_{1}$ and $m_{2}$ to fulfill $m_{1}+ m_{2} = M_{2}$, and the second sum runs over all permutations $\mathcal{P}$ of the set $\{\alpha_{k}\} = \{\alpha_{k_{1}},\ldots,\alpha_{k_{M_{2}}} \}$ containing $m_{1}$ times the phase prefactor $\alpha_{1}=0$ and $m_{2}$ times the phase prefactor $\alpha_{2}=1$. However, as we have seen in Sects.~\ref{sec:Setup1} and \ref{sec:Setup2}, this specific sum with $M_{2}$ detectors at the MP can only be unequal to zero if $m_{1}=M_{2}$ and $m_{2}=0$, or $m_{1}=0$ and $m_{2}=M_{2}$, i.e., all photon annihilation operators have to act on the same source. By use of Eqs.~(\ref{eq:sum_detectorphases0}) and (\ref{eq:sum_detectorphases}) we thus obtain 
\begin{align}
	\prod_{j=2}^{M_{2}+1} \hat{E}^{(+)}(\delta_{j}) = \hat{a}_{1}^{M_{2}} + (-1)^{M_{2}-1} \hat{a}_{2}^{M_{2}}  \, .
\label{eq:FieldMP2}
\end{align}
Plugging Eq.~(\ref{eq:FieldMP2}) into Eq.~(\ref{eq:StateMP}) the projected state $\tilde{\rho}_{{}_{2\, \text{TLS}}}^{(M_{2})}$ of the two TLS after $M_{2}$ photons have been detected at the MP takes the form
\begin{align}
	\tilde{\rho}_{{}_{2\, \text{TLS}}}^{(M_{2})} &= \hat{a}_{1}^{M_{2}} \rho_{{}_{2\, \text{TLS}}}\, \hat{a}_{1}^{\dagger \, M_{2}} + \hat{a}_{2}^{M_{2}} \rho_{{}_{2\, \text{TLS}}}\, \hat{a}_{2}^{\dagger \, M_{2}} \nonumber \\
	& \phantom{=} + (-1)^{M_{2}-1} \left[ \hat{a}_{1}^{M_{2}} \rho_{{}_{2\, \text{TLS}}}\, \hat{a}_{2}^{\dagger \, M_{2}} + \hat{a}_{2}^{M_{2}} \rho_{{}_{2\, \text{TLS}}}\, \hat{a}_{1}^{\dagger \, M_{2}} \right] \, .
\label{eq:StateMP2TLS}
\end{align}
For $M_{2}=1$ this state becomes identical to the superradiant case discussed in \cite{Bhatti(2018)}.
For $M_{2} \geq 2$ we see that besides the diagonal terms characterizing the thermal nature of the density matrix only two further nondiagonal terms appear.
Performing a subsequent $G^{(M_1)}$ measurement with $M_1=M_2$ moving detectors, we find that exactly these two nondiagonal terms are responsible for the $M_{2}$-photon interferences leading to the N00N-like modulation with $N=M_{2}$ (see also Figs.~\ref{fig:N00N_quantum_paths_setup1} and \ref{fig:N00N_quantum_paths_setup2}). Consequently, the state given in Eq.~(\ref{eq:StateMP2TLS}) can be interpreted as the classical analog to the quantum mechanical N00N-state.


\section{Experimental results}
\label{sec:Experiment}

\begin{figure}[b]
	\centering
		\includegraphics[width=1.\columnwidth]{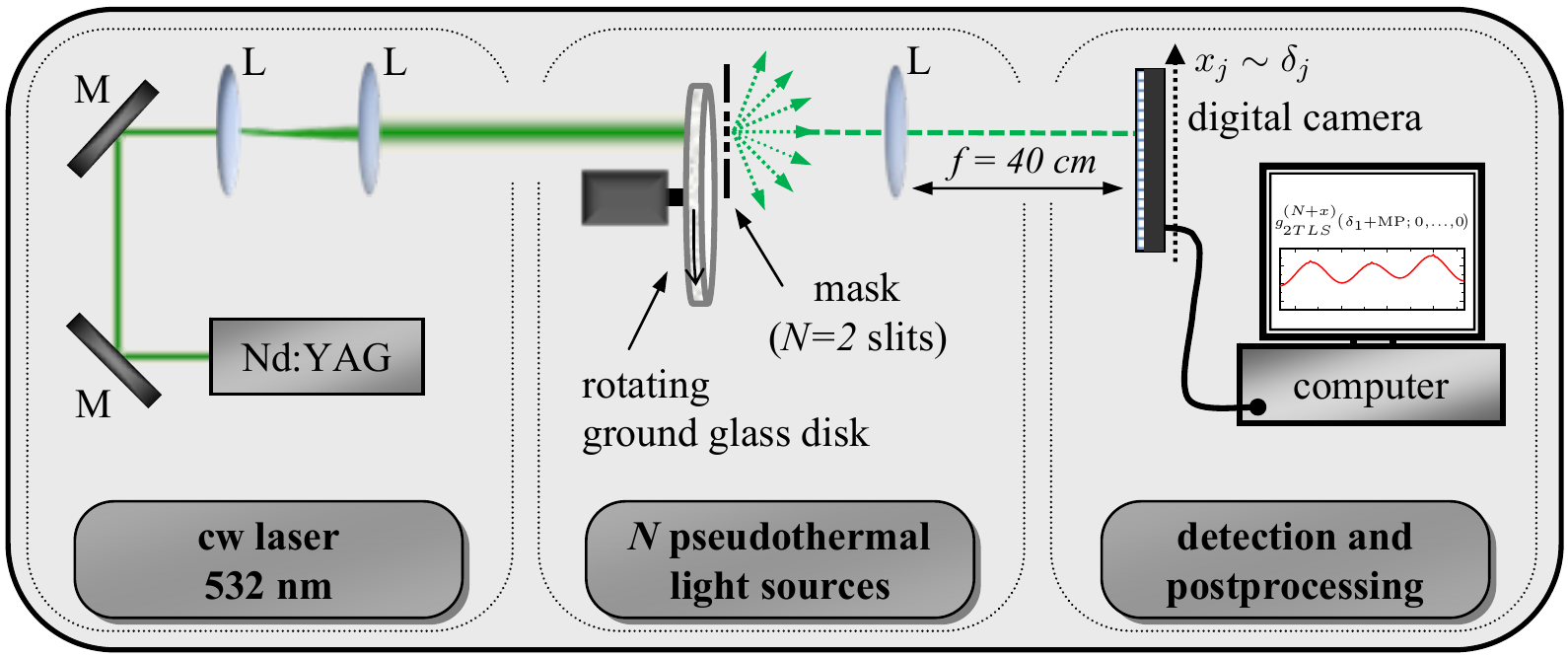}
	\caption{Experimental setup to measure $\bar{g}_{N\,\text{TLS}}^{(M)}$ with two pseudothermal light sources. For experimental details see the text and Ref.~\cite{Oppel(2014)}. M, Mirror. L, Lens.}
\label{fig:Setup}
\end{figure}

To verify the theoretical predictions, in particular of setup 2 presented in Sect.~\ref{sec:Setup2}, we evaluated higher-order intensity correlation functions of light fields originating from two independent pseudothermal light sources. To that aim  we used linearly polarized light from a frequency-doubled Nd:YAG laser at $\lambda = 532$~nm scattered from a rotating ground glass disc to produce a spatially and temporally varying speckle field with thermal statistics \cite{Estes(1971)}. The rotation speed of the ground glass disc was chosen to create a second-order coherence time of $\tau_{c} \approx 50$~ms. We utilzed this light to illuminate a double slit mask such that the slits acted as two statistically independent pseudothermal light sources.  
The two identical slits of width $a = 25$ $\mu$m and separation $d = 200$ $\mu$m produced sufficient light to work in the high intensity regime. With this setup we recorded a series of $\sim 10^{4}$ speckle images 
in the far field of the mask utilizing a conventional digital camera with an integration time $\tau_{i} \approx 1\, \text{ms} \ll \tau_{c}$ \cite{Oppel(2014)}. From these images we evaluated the normalized $M$th-order correlation functions $\bar{g}_{2\,\text{TLS}}^{(M)} = G_{2\,\text{TLS}}^{(M)}/\text{max}\{G_{2\,\text{TLS}}^{(M)} \}$ by cross correlating the intensity grey values of $M_{2}$ fixed pixels at the $\text{MP}$ with $M_{1}$ moving pixels at $\delta_{1}$. 

\begin{figure}[t]
	\centering
		\includegraphics[width=1.\columnwidth]{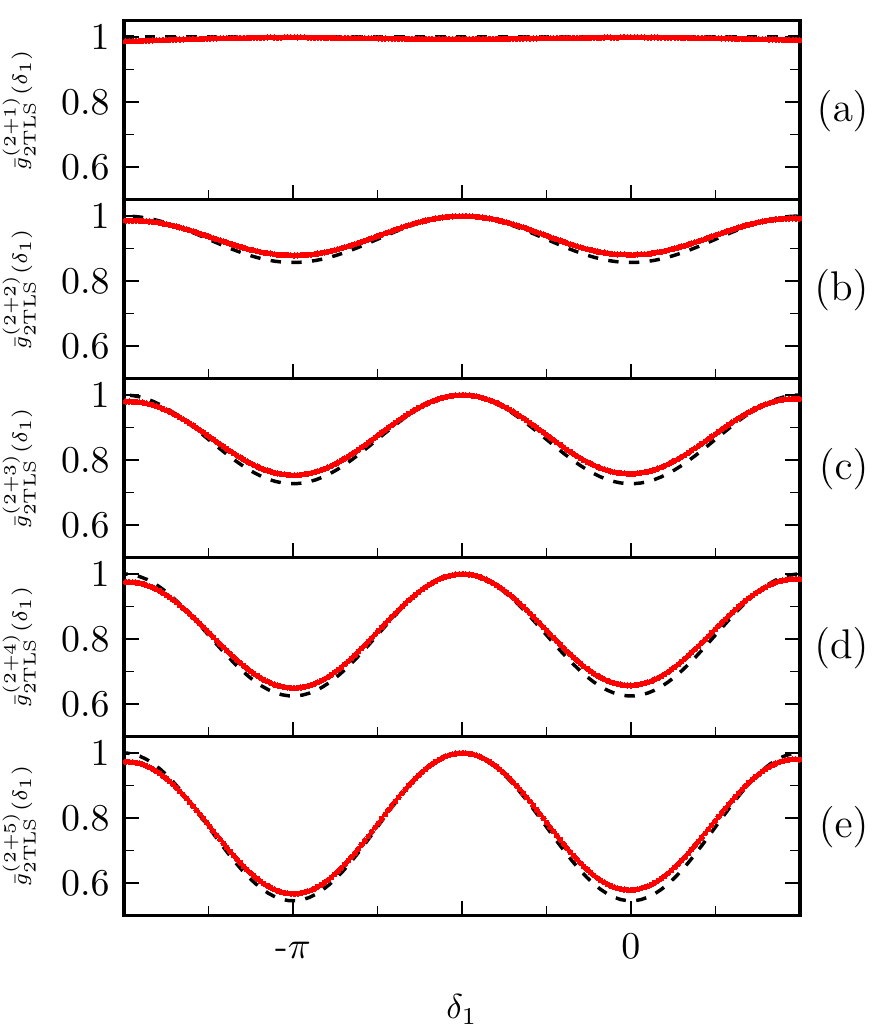}
	\caption{Experimental results for $M_2=2$ detectors placed at the magic positions (MP), while $M_1=1,2,3,4,5$ detectors are placed at the position $\delta_1$.}
	\label{fig:g(2+x)}
\end{figure}

Fig.~\ref{fig:g(2+x)} displays the experimental results, i.e., the correlation functions $\bar{g}_{2\,\text{TLS}}^{(2+M_{1})}(\delta_{1},\ldots,\delta_{1};0,\pi)$ for $M_{1}=1,\ldots,5$ (red dotted curves), together with the theoretical predictions (black dashed curves) [see Eqs.~(\ref{eq:GNgeneralform_2}) and (\ref{eq:GNgeneralform_2_vis})].
In the case of $M_{1}=1 < M_{2}=2$ (Fig.~\ref{fig:g(2+x)}a) no modulation is visible, whereby a clear modulation $\cos(2\delta_{1})$ appears for $M_{1}=2$ (Fig.~\ref{fig:g(2+x)}b). The cosine modulation increases in visibility when increasing $M_{1}=3,4,5$ (Fig.~\ref{fig:g(2+x)}c-e), in excellent agreement with the theory.

\begin{figure}[t]
	\centering
		\includegraphics[width=1.\columnwidth]{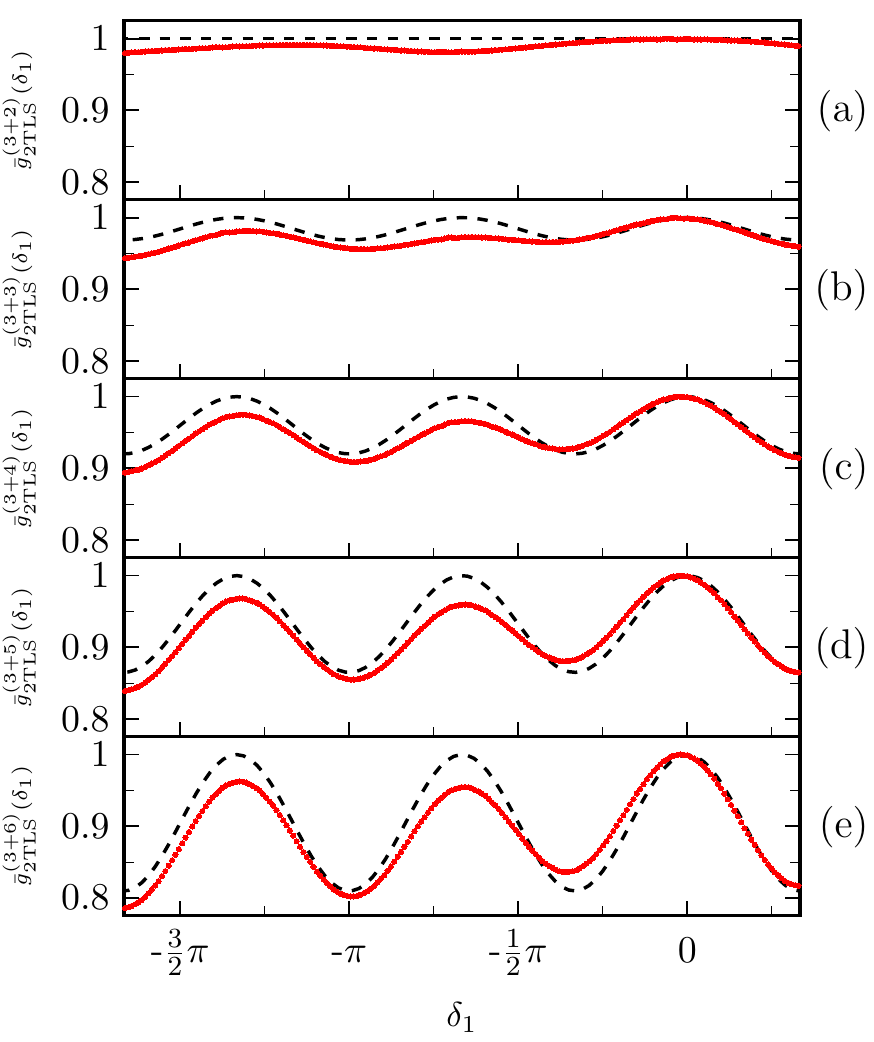}
	\caption{Experimental results for $M_2=3$ detectors placed at the magic positions (MP), while $M_1=2,3,4,5,6$ detectors are placed at the position $\delta_1$.}
	\label{fig:g(3+x)}
\end{figure}

Fig.~\ref{fig:g(3+x)} shows the experimental results for \linebreak[4] $\bar{g}_{2\,\text{TLS}}^{(3+M_{1})}(\delta_{1},\ldots,\delta_{1};0,2\pi/3,4\pi/3)$ with $M_{1}=2,\ldots,6$. \linebreak[4] Again, for $M_{1}=2 < M_{2}=3$,
no modulation arises [see Fig.~\ref{fig:g(3+x)}a].
Only for $M_{1}\geq 3$ moving detectors the predicted modulation $\cos(3\delta_{1})$ appears, where the visibility increases anew with increasing the number of moving detectors (see Fig.~\ref{fig:g(3+x)}b-e). Again, the measurements are in good agreement with the theoretical predictions. The small deviations can be explained by the finite extent of the slits, leading to a more involved envelope function, and the discrete  pixel size of the digital camera, impeding the precise choice of the MP (for a detailed discussion see Refs.~\cite{Classen(2016),Schneider(2018)}).


\section{Discussion}
\label{sec:Discussion}

For first-order coherent diffraction or second-order HBT intensity interferometry a two-emitter geometry as considered in this paper yields in the Fourier plane the modulation $\cos(\delta_1)$ \cite{Loudon(1968)}.  
According to Abbe`s resolution limit of classical optics \cite{Abbe(1873),MANDEL(1995)}, in order to resolve the two emitters, two adjacent diffraction maxima need to be captured in the Fourier plane by the numerical aperture of the imaging device, which for a modulation $\cos(\delta_1)$ corresponds to $\delta_1 \in [0,2\pi]$ (in phase units). 
By contrast, the detection schemes discussed in Sects.~\ref{sec:Setup1} and \ref{sec:Setup2} generate for the same emitter geometry  in the Fourier
plane modulations of the form $\cos(M_2 \delta_1)$ [see Eq.~(\ref{eq:GNgeneralform})].
According to the classical resolution limit, 
the faster modulations 
reduce the required numerical aperture by the factor $M_{2}$, i.e., the resolution power is $M_{2}$-fold enhanced.

In previous approaches we discussed detector configurations enabling the filtering of the fastest modulations that are contained in the Fourier spectrum of the source geometry, equally leading to sub-Abbe resolution \cite{Oppel(2012),Classen(2016),Schneider(2018)}. By contrast, in the present scheme, we do not rely on a filtering mechanism but produce true N00N-like modulations with an effectively $M_{2}$-fold reduced de Broglie wavelength. This is due to the isolation of the M-photon quantum paths discussed in Sect.~\ref{sec:Setup2} where each TLS effectively emits $M_{2}$ photons that collectively propagate to the two different sets of detectors.

The approach bears the potential to increase the resolution in lens-less far-field imaging, relevant for, e.g., astronomy or other fields of research where lenses cannot be employed. For example, a double-star system could be imaged with two-fold enhanced resolution and a visibility $\mathcal{V}_{2+M_{1}}$ (approaching unity for increased $M_1$) when measuring  $G_{2\,\text{TLS}}^{(2+M_{1})}(\delta_{1},\hdots,\delta_{1};0,\pi)$ compared to a simple HBT measurement $G_{2\,\text{TLS}}^{(2)}(\delta_{1},0)$ with a maximum visibility $\mathcal{V}_{2} = 1/3$ [see Eqs.~(\ref{eq:GNgeneralform_2}), (\ref{eq:GNgeneralform_2_vis}) and (\ref{eq:N00NGN}), respectively]. In the future, the approach might be extended to more than two emitters by carefully tailoring appropriate detection schemes so that quantum paths are isolated such that photons from each emitter collectively propagate toward different sets of detectors.

We finally note that unlike previous implementations that produce N00N-like modulations in first-order intensity correlations via appropriate superpositions of coherent light fields \cite{Resch(2007),Kothe(2010),Shabbir(2013)} we consider in this paper self-luminous and statistically independent incoherent sources. 
For these sources the first-order intensity correlation function yields a constant,  
and it is only the spatial intensity correlations of higher order that reveal structural information about the extension and location of the sources.

\section{Conclusion}
\label{sec:Conclusion}

In conclusion we have shown in this paper that for a system of two TLS N00N-like interferences of arbitrary order $N=M_2$ can be produced via measurement of higher-order intensity correlation functions. In particular, we demonstrated that by recording $M_{2}$ photons at fixed so-called magic positions [see Eq.~(\ref{eq:MagicPositionsM})] 
the two TLS are projected into a highly correlated \textit{classical N00N-state} [see Eq.~(\ref{eq:StateMP2TLS})] displaying a similar structure as its quantum mechanical counterpart with identical quantum path interferences. Subsequently measuring coincidentally $M_{1}\geq M_{2}$ photons with $M_{1}$ moving detectors, the classical N00N state displays N00N-like interferences of order $N = M_2$. 
Two different schemes for such $M_{1} + M_{2}=M$th-order correlation measurement were discussed in Sects.~\ref{sec:Setup1} and \ref{sec:Setup2} and corresponding experimental results were presented in Sect.~\ref{sec:Experiment} for $N=2$, and $N=3$, in excellent agreement with the theoretical predictions.

The algorithms presented in this paper could be of interest for any lens-less far-field imaging scheme, e.g., 
in astronomy where recent investigations of HBT measurements revived the field of intensity interferometry for the observation of distant stars \cite{Dravins(2013),Trippe(2014),Tan(2016),Guerin(2017),Pilyavsky(2017),Rivet(2018),Guerin(2018),Matthews(2018)}. In particular, the possibility to increase the resolution in intensity interferometry by use of higher-order correlation measurements with detectors at specific positions might be of interest in this context.

We close in noting that setup 2 discussed in Sect.~\ref{sec:Setup2} could also be of use for lithography. By sending the light of two TLS on a beam splitter and performing a projective measurement of $M_{2}$ photons in one of the two output ports of the beam splitter would lead to a highly modulated $M_{1}$-photon interference pattern in the other output port. Letting the modulated light field in the second output port impinge on a $M_1$-photon resist (with $M_{1}\geq M_{2}$) would produce a highly modulated N00N-state interference pattern with $N = M_2$ [see Eq.~(\ref{eq:GNgeneralform_2})]. 

\section{Acknowledgement}

D.B. and J.v.Z. thank C.~A. Str\"ohlein and G.~S. Agarwal for helpful comments and very fruitful discussions.
The authors gratefully acknowledge funding by the Erlangen Graduate School in Advanced Optical Technologies (SAOT) by the German Research Foundation (DFG) in the framework of the German excellence initiative. D.B. gratefully acknowledges financial support by the Cusanuswerk, 
Bisch\"ofliche Studienf\"orderung.


\newpage

%
%

%
%

\end{document}